\documentclass[journal=jacsat,manuscript=article]{achemso}
\usepackage{chemformula} 
\usepackage[T1]{fontenc} 
\usepackage{graphicx}
\usepackage{dcolumn}
\usepackage{natbib}
\usepackage{amsmath,amsfonts,amssymb,bm}
\usepackage[version=3]{mhchem} 
\usepackage{xcolor}
\usepackage{amsmath}
\usepackage{placeins}
\usepackage{float,afterpage} 

\author{Haoran Liu}
\affiliation[East China Normal University]{State Key Laboratory of Precision Spectroscopy Science and Technology, East China Normal University,200241 Shanghai, China}

\author{Zihe Jiang}
\affiliation[East China Normal University]{State Key Laboratory of Precision Spectroscopy Science and Technology, East China Normal University,200241 Shanghai, China}

\author{Zhiwei Hu}
\affiliation[East China Normal University]{State Key Laboratory of Precision Spectroscopy Science and Technology, East China Normal University,200241 Shanghai, China}

\author{Banghuan Zhang}
\affiliation[East China Normal University]{State Key Laboratory of Precision Spectroscopy Science and Technology, East China Normal University,200241 Shanghai, China}

\author{Tao He}
\affiliation[East China Normal University]{State Key Laboratory of Precision Spectroscopy Science and Technology, East China Normal University,200241 Shanghai, China}

\author{Xiaohui Dong}
\affiliation[East China Normal University]{State Key Laboratory of Precision Spectroscopy Science and Technology, East China Normal University,200241 Shanghai, China}

\author{Chaowei Sun}
\affiliation[Henan Academy of Sciences]
{Institute of laser Manufacturing, Henan Academy of Sciences, Zhengzhou, China.}

\author{Jun Tian}
\affiliation[Fudan University]{Department of Chemistry, State Key Laboratory of Porous Materials for Separation and Conversion, Fudan University, 200438 Shanghai, China}

\author{Wei Jiang}
\affiliation[Henan Academy of Sciences]
{Institute of laser Manufacturing, Henan Academy of Sciences, Zhengzhou, China.}

\author{Huatian Hu}
\email{huatian.hu@iit.it}
\affiliation[Istituto Italiano di Tecnologia]{Istituto Italiano di Tecnologia, Center for Biomolecular Nanotechnologies, Via Barsanti 14, 73010 Arnesano, Italy}

\author{Wen Chen}
\email{wchen@lps.ecnu.edu.cn}
\affiliation[East China Normal University]{State Key Laboratory of Precision Spectroscopy Science and Technology, East China Normal University,200241 Shanghai, China}

\author{Hongxing Xu}
\email{hxxu@hnas.ac.cn}
\affiliation[Henan Academy of Sciences]
{Institute of laser Manufacturing, Henan Academy of Sciences, Zhengzhou, China.}
\alsoaffiliation[East China Normal University]{State Key Laboratory of Precision Spectroscopy Science and Technology, East China Normal University,200241 Shanghai, China}

\title[An \textsf{achemso} demo]
  {Spatiotemporal Raman Probing of Molecular Transport in sub-2-nm Plasmonic Quasi-2D Nanochannels }

\abbreviations{IR,NMR,UV}

\begin{document}


\begin{abstract}

\noindent  
Capturing molecular dynamics in nanoconfined channels with high spatiotemporal resolution is a key challenge in nanoscience, crucial for advancing catalysis, energy conversion, and molecular sensing. Bottom-up ultrathin plasmonic nanogaps, such as nanoparticle-on-mirror (NPoM) structures, are ideal for ultrasensitive probing due to their extreme light confinement, but their perceived sealed geometry has cast doubt on the existence of accessible transport pathways.
Here, counterintuitively, we demonstrate that ubiquitous ligand-capped NPoM-type nanogaps can form a natural quasi-two-dimensional nanochannel, supporting molecular transport over unprecedented length scales ($\gtrsim5$ $\mu$m) with an extreme aspect ratio ($>10^3$). 
Using wavelength-multiplexed Raman spectroscopy, we resolve the underlying centripetal infiltration pathway with a spatial resolving power of $\sim$20 nm. 
This redefines the NPoM architecture as a sensitive, \textit{in-situ}, all-in-one "transport-and-probe" platform,
enabling real-time, reusable monitoring of analyte with $\sim$10$^{-11}$ M. 
This work establishes a versatile new platform for advancing super-resolved \textit{in-situ} molecular sensing, nanoscale physicochemical studies, and on-chip nanophotofluidics.

\end{abstract}

\section{Introduction}
Probing and understanding molecular dynamics within nanoconfined environments is fundamental to surface physical chemistry and nanoscience. 
\cite{leeIntegratingValidatingExpanding2024,
shenArtificialChannelsConfined2021,
spitzbergPlasmonicNanoporeBiosensorsSuperior2019,
garoliPlasmonicNanoporesSingleMolecule2019}
Confinement within low-dimensional geometries—such as 0D nanopores, 
\cite{liuMixedMatrixFormulations2018,
spitzbergPlasmonicNanoporeBiosensorsSuperior2019,
garoliPlasmonicNanoporesSingleMolecule2019}
1D nanotubes, 
\cite{tunuguntlaEnhancedWaterPermeability2017}
and 2D nanochannels
\cite{keerthiBallisticMolecularTransport2018,
zhangControllableIonTransport2019}
—imposes physical constraints that radically alter molecular transport, reactivity, and selectivity compared to bulk systems, giving rise to unique applications ranging from molecular sieving to ultrafast and dynamically tunable transport. 
\cite{lozada-hidalgoSievingHydrogenIsotopes2016,
tunuguntlaEnhancedWaterPermeability2017, 
zhouElectricallyControlledWater2018,
liuMixedMatrixFormulations2018,
keerthiBallisticMolecularTransport2018,
zhangControllableIonTransport2019}
While these phenomena can be characterized by various means including electrical readouts 
\cite{lozada-hidalgoSievingHydrogenIsotopes2016,
tunuguntlaEnhancedWaterPermeability2017,
zhouElectricallyControlledWater2018}
and structural imaging,
\cite{keerthiBallisticMolecularTransport2018}
optical methods offer particularly powerful, non-invasive routes for tracking molecular behavior in real time. 
\cite{spitzbergPlasmonicNanoporeBiosensorsSuperior2019,
garoliPlasmonicNanoporesSingleMolecule2019}
Yet, conventional optics are restricted by the diffraction limit and intrinsically weak light-matter interactions, creating a critical need for novel approaches that provide label-free, single-molecule sensitivity with high spatiotemporal resolution. 

Plasmonic nanostructures can overcome this challenge by concentrating light into deep-subwavelength volumes, thereby greatly enhancing light–matter interactions. 
Recent advances in plasmonics have ushered in an era of \textit{extreme nanophotonics}, 
\cite{liPlasmonicParticleonfilmNanocavities2018,
baumbergExtremeNanophotonicsUltrathin2019,
wang2020fundamental,
liBoostingLightMatterInteractions2024a}
where metallic nanogaps can trap light into 1 nm$^3$ volume, achieving atomic-scale sensitivity,
\cite{benzSinglemoleculeOptomechanicsPicocavities2016a,
chenIntrinsicLuminescenceBlinking2021}
opening doors for vast novel applications such as single-molecular science,
\cite{xuSpectroscopySingleHemoglobin1999a,
zhangChemicalMappingSingle2013,
choiMetalCatalyzedChemicalReaction2016,
kimSmartSERSHot2018,
langerPresentFutureSurfaceEnhanced2020}
surface chemistry. 
\cite{choiMetalCatalyzedChemicalReaction2016,
wangSituRamanSpectroscopy2021a,
oksenbergEnergyresolvedPlasmonicChemistry2021a,
zhanPlasmonicNanoreactorsRegulating2021,
huAlchemicallyglazedPlasmonicNanocavities2025}
The integration of plasmonics with nanofluidics has yielded powerful analytical platforms. 
\cite{garoliPlasmonicNanoporesSingleMolecule2019,
spitzbergPlasmonicNanoporeBiosensorsSuperior2019}
A prominent example is the plasmonic nanopore, typically fabricated via top-down methods, which funnels analytes through electrophoresis \cite{chenHighSpatialResolution2018b} or optical trapping \cite{belkinPlasmonicNanoporesTrapping2015}
for applications like DNA and protein analysis. 
\cite{
belkinPlasmonicNanoporesTrapping2015,
assadLightEnhancingPlasmonicNanoporeBiosensor2017,
shiActiveDeliverySingle2018,
chenHighSpatialResolution2018b,
zhaoPlasmonicBowlShapedNanopore2023,
zhouSingleMoleculeProtein2023} 
These systems enable real-time characterization through signals such as fluorescence enhancement,
\cite{assadLightEnhancingPlasmonicNanoporeBiosensor2017}
resonance shift 
\cite{shiActiveDeliverySingle2018},
or surface-enhanced Raman spectroscopy (SERS)
\cite{chenHighSpatialResolution2018b,
zhouSingleMoleculeProtein2023}.
However, the 0D nature of nanopores inherently limits analyte dwell time, which hinders the detection of weak or specific signals as well as the observation of subtle or slow dynamic processes.
\cite{garoliPlasmonicNanoporesSingleMolecule2019,
spitzbergPlasmonicNanoporeBiosensorsSuperior2019}
Furthermore, the top-down fabrication of reliable, sub-5-nm gaps remains a considerable and costly challenge—a limitation shared by other top-down nanogap configurations which often suffer from poor geometric control. 
\cite{luoScalableFabricationMetallic2021}
In contrast, 2D nanochannels offer an extended spatial dimension, providing a much longer observation window ideal for studying molecular infiltration and interaction dynamics. 
However, achieving high-spatiotemporal-resolution probing of molecular transport in such 2D plasmonic nanochannels remains challenging.

Among plasmonic architectures, the nanoparticle-on-mirror (NPoM) geometry stands out as a highly tunable, high-quality-factor, and reproducible nanoplatform. 
\cite{liPlasmonicParticleonfilmNanocavities2018,
baumbergExtremeNanophotonicsUltrathin2019,
chenProbingLimitsPlasmonic2018}
Formed through simple, high-yield bottom-up assembly, it offers well-defined and consistent gap dimensions toward a sub-nanometer scale that supports huge field enhancement. Intuitively, the insulating layer that mechanically supports the nanoparticle in this vertically assembled metal–insulator–metal nanogap \cite{baumbergExtremeNanophotonicsUltrathin2019} would appear to block and seal the 2D nanochannel formed between the two metallic surfaces in the gap.
Consequently, their application has been predominantly focused on probing pre-embedded analytes—such as biomolecules, self-assembled monolayers or 2D materials
\cite{akselrodProbingMechanismsLarge2014a,
choiMetalCatalyzedChemicalReaction2016,
chenProbingLimitsPlasmonic2018,
kimSmartSERSHot2018,
chikkaraddyMappingNanoscaleHotspots2018b,
liObservationInhomogeneousPlasmonic2020}
—or has been limited to leveraging only the edge regions of the nanogap for post-sensing,
\cite{imSelfAssembledPlasmonicNanoring2013a,chenLargeScaleHotSpot2015}
rather than exploiting the central hotspot as an open platform for dynamic sensing.
\cite{taylor2017single}
Parallel efforts to engineer \textit{open} bottom-up nanogaps for molecular access and transport have been pursued 
\cite{chenProbingSubpicometerVertical2018,
kimSmartSERSHot2018,
liuMetalOrganicFrameworkEnabled2024}, 
for example, through host–guest chemistry 
\cite{kimSmartSERSHot2018}
and metal–organic frameworks (MOFs)
\cite{liuMetalOrganicFrameworkEnabled2024}. 
Additional post-fabrication methods such as surface cleaning,
\cite{ansarRemovalMolecularAdsorbates2013,zhongRemovalResidualPolyvinylpyrrolidone2019}
ligand exchange
\cite{moranReplacementPolyvinylPyrrolidone2011,
villarrealNanoscaleSurfaceCurvature2017,
pereraFacileDisplacementCitrate2018,
zhouEnablingCompleteLigand2018,
ahnComparativeStudyAdsorption2019,
oksenbergEnergyresolvedPlasmonicChemistry2021a,
lapresta-fernandezSiteselectiveSurfaceEnhanced2022}
and electrochemical cycling 
\cite{huangCleanModifiedSubstrates2011,
viehrigQuantitativeSERSAssay2020,
sibug-torresSituElectrochemicalRegeneration2024}
have been employed to reuse the plasmonic surfaces or gaps.
However, these studies have largely focused on the ensemble-averaged effects of static adsorption or \textit{ex situ} detection. 
The real-time, spatially resolved observation of molecular infiltration dynamics within an individual gap has remained unaddressed. 
This creates a clear division: top-down structures with open channels allow post-fabrication sensing but suffer from limited performance, while bottom-up systems like NPoM offer superior optical properties but are often treated as closed or their internal analyte dynamics remain a black box. 

\begin{figure}[!ht]
  \centering
  \captionsetup{font=small}
  \includegraphics[width=0.95\textwidth]{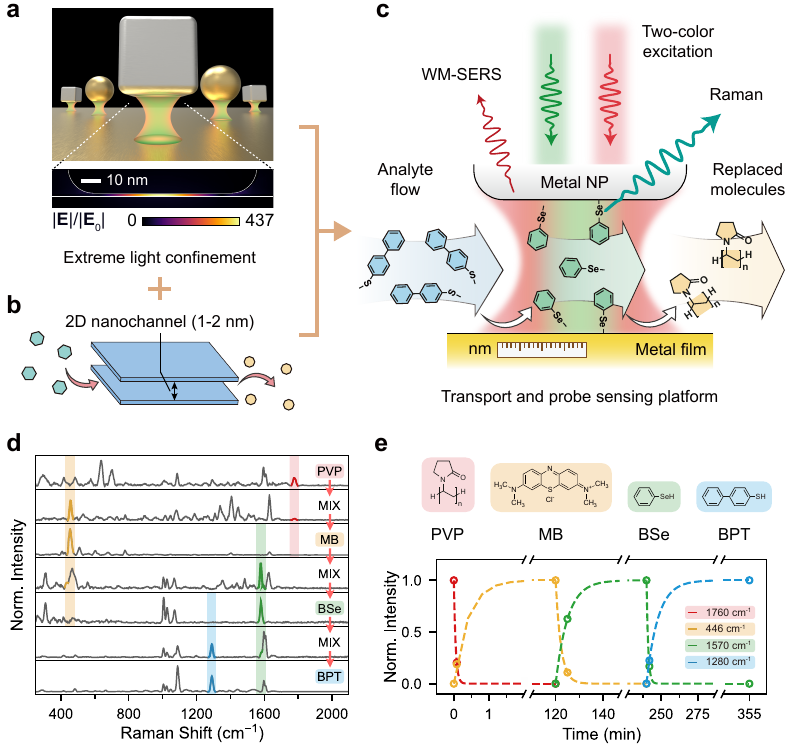}
  \caption{\textbf{Sequential molecule-exchange SERS sensing in NPoM plasmonic nanochannels.}     
    (\textbf{a}) Top panel: 3D schematics of metal nanoparticles (nanosphere, nanocube, etc.) placed over an metal mirror to form well-defined plasmonic nanogaps (top panel); Bottom panel: Simulated field distribution of a nanocube-on-mirror (NCoM) with 1 nm gap, exhibiting pronounced localized field enhancement ($|\mathbf{E}| / |\mathbf{E}_0| =437$). 
    (\textbf{b}) Schematics of a molecule exchanges in a 2D molecular nanochannel. 
    (\textbf{c}) Schematic illustration of super-resolved wavelength-multiplexed (WM) SERS spectroscopy for sensing of sequential molecular exchange dynamics in 1-2 nm thick 2D plasmonic nanochannels. 
    (\textbf{d}) Normalized SERS spectra (785 nm excitation) of a single PVP-capped NCoM before and after the sequential immersion in the solution of MB $\rightarrow$ BSe $\rightarrow$ BPT (from top to bottom). 
    The characteristic peaks of each molecule are highlighted in red, yellow, green, and blue area, respectively.  
    (\textbf{e}) Molecular structures of PVP, MB, BSe and BPT (top panel), and their corresponding characteristic SERS peak intensities (colored open circles, extracted from \textbf{d}) as a function of immersion duration, where the dash lines are the fitting results by Langmuir kinetic equation (See details in Supplementary Section S8).
    }
  \label{fig:1}
\end{figure}

In this work, counterintuitively, we demonstrate that the simplest and ubiquitous molecular-spaced plasmonic nanogaps naturally host quasi-2D nanochannels that can support molecular transport and interactions (Fig.~\ref{fig:1}a-c), which can be spatiotemporally resolved using wavelength-multiplexed Raman spectroscopy (WM-SERS).
\cite{wangProbingLocationHot2014,griffiths_locating_2021}
By harnessing the spatially complementary near-fields of distinct plasmonic modes, our WM-SERS technique achieves a spatial resolving power of 20 nm.
In the end, we show how the NPoM configuration can be redefined as an all-in-one transport-and-probe nanoplatform, where nanoconfined optical field overlaps with a natural functional nanochannel in the nanogap to enable and probe \textit{in situ} molecular science. 
This natural quasi-2D nanochannel enables nanoconfined analyte transport via molecular exchange dynamics, for which we establish a mechanical model revealed by WM-SERS: target molecules progressively diffuse from the nanogaps' edge to central regions, achieving maximal SERS via a ligand-exchange mechanism, whose efficiency is dependent on molecular binding affinities.

Furthermore, this molecular transport process can occur over unprecedented length scales ($\gtrsim5$ $\mu$m) within tightly confined ($\sim2$ nm) nanochannels, with an extreme aspect ratio of $>10^3$, whose trajectory can be real-time traced by SERS mapping on microplate-on-foil (MPoF) nanogaps. 
Additionally, we further demonstrate that this platform enables real-time monitoring of nanoconfined molecular transport with microfluidic on-chip integration. Individual NPoM channels can achieve high sensitivity (10$^{-11}$M) for nonresonant small analytes toward single-molecule level using state-of-the-art digital SERS techniques.
Our work introduces a versatile platform that merges extreme optical confinement with active mass transport, facilitating novel applications for super-resolved molecular sieving and sensing, nanoreactors, confined-phase chemistry, and nanophotofluidics. 

\section{Results}
\subsection{Sequential molecule-exchange in a single plasmonic nanogap.}
We begin by monitoring molecular exchange dynamics inside the well-defined 1-2 nm nanogap through SERS measurements to validate its role as a 2D nanochannel. Fabrication of plasmonic vertical nanogaps was achieved via direct deposition of chemically synthesized colloidal nanoparticles onto a gold film (See details in Supplementary Section S1). As shown in Fig.~\ref{fig:1}c, the nanogap acts as an optical nanocavity providing extreme light confinement (enhancement compared with free-space incidence $|\mathbf{E}|/|\mathbf{E}_0|\simeq 437$, Fig.~\ref{fig:1}a), while simultaneously serving as a quasi-2D molecular nanochannel that enables molecular transports and interactions inside (Fig.~\ref{fig:1}b). The surfactant ligand layers inherent to these nanoparticles—commonly citrate, polyvinylpyrrolidone (PVP), or cetyltrimethylammonium chloride (CTAC)—function as 1-3 nm thick molecular spacers, creating well-defined nanochannels conducive to controlled analyte infiltration and exchange in subsequent sensing applications.
To verify this infiltration and exchange mechanism in the nanochannel, we will use single-color SERS to probe the molecular components, and two-color excitation WM-SERS to spatiotemporally resolve their exchange dynamics within an NPoM. 

We first employed a 70-nm-diameter PVP-capped silver nanocube placed on a mirror (NCoM). This system underwent sequential incubation cycles in solutions of methylene blue (MB), benzenethiol (BSe), and biphenyl-4-thiol (BPT), utilizing the PVP layer as the initial molecular spacer. 
Before and after each incubation step, single-nanoparticle SERS was performed using 785 nm laser excitation to monitor the molecular exchange dynamics in time within the same nanogap (nanochannel). The resultant spectral evolution of the complete replacement sequence (PVP → MB → BSe → BPT) is shown in Fig.~\ref{fig:1}d.
Upon immersion in MB for merely 5 seconds, characteristic PVP Raman peaks (red shade) were partially replaced by mixed spectral signatures of both PVP and MB. Extended incubation (2 hours) led to near-complete molecular exchange, with SERS spectra dominated exclusively by MB vibrational modes (yellow shade). Subsequent incubations in BSe and BPT solutions exhibited similar spectral changes, confirming efficient replacement of the pre-adsorbed molecular layers inside and outside the nanogaps. 
Figure \ref{fig:1}e quantifies this multistep replacement dynamics, showing the normalized intensities of representative vibrational bands (demonstrated by colored regions in Fig. \ref{fig:1}d) against incubation time. 
Taking the lifecycle of the MB molecule as an example, the curve traces the complete five-stage process (the five yellow open circles from left to right in Fig.~\ref{fig:1}e): (i) its initial absence, (ii) the partial and then (iii) saturated replacement of the preceding PVP layer, (iv) its subsequent partial replacement by the incoming BSe molecules, and finally (v) its complete replacement from the nanochannel. This analysis provides a qualitative insight into the continuous molecular exchange occurring within the nanochannel.

To describe the temporal evolution of the molecular exchange, we applied a simplified Langmuir kinetic model to our experimental data (details in Supplementary Section S8).
\cite{tukovaShapeInducedVariationsAromatic2023} 
The fits, shown as dashed lines in Fig.~\ref{fig:1}e, effectively model the observed saturation dynamics. 
The extracted phenomenological rate parameters reveal a clear trend: the replacement of the weakly-bound PVP spacer is exceptionally rapid, whereas the final exchange between BSe and BPT, two molecules with similar thiol-gold bonds, is the slowest step. 
This kinetic analysis provides a qualitative insight into the exchange process, highlighting that the platform can distinguish between analytes based on their apparent dynamic binding properties at the nanoscale.
This analysis validates the proposed mechanism and highlights the platform's capability to differentiate between analytes based on their dynamic binding properties at the nanoscale.
Notably, this molecular transport is not strictly unidirectional or dictated solely by binding affinity. 
We found that even molecules with weaker binding energies, such as MB, can progressively replace the strongly-bound BPT in the nanogap, albeit at a slower rate (see Supplementary Fig. S12). 
This demonstrates the bidirectional molecular transport capability of the nanochannel, underscoring its nature as a truly open and versatile platform.


\subsection{Spatiotemporally resolved molecular infiltration and exchange dynamics in single nanogaps.}
\begin{figure}[!ht]
  \centering
  \includegraphics[width=1\textwidth]{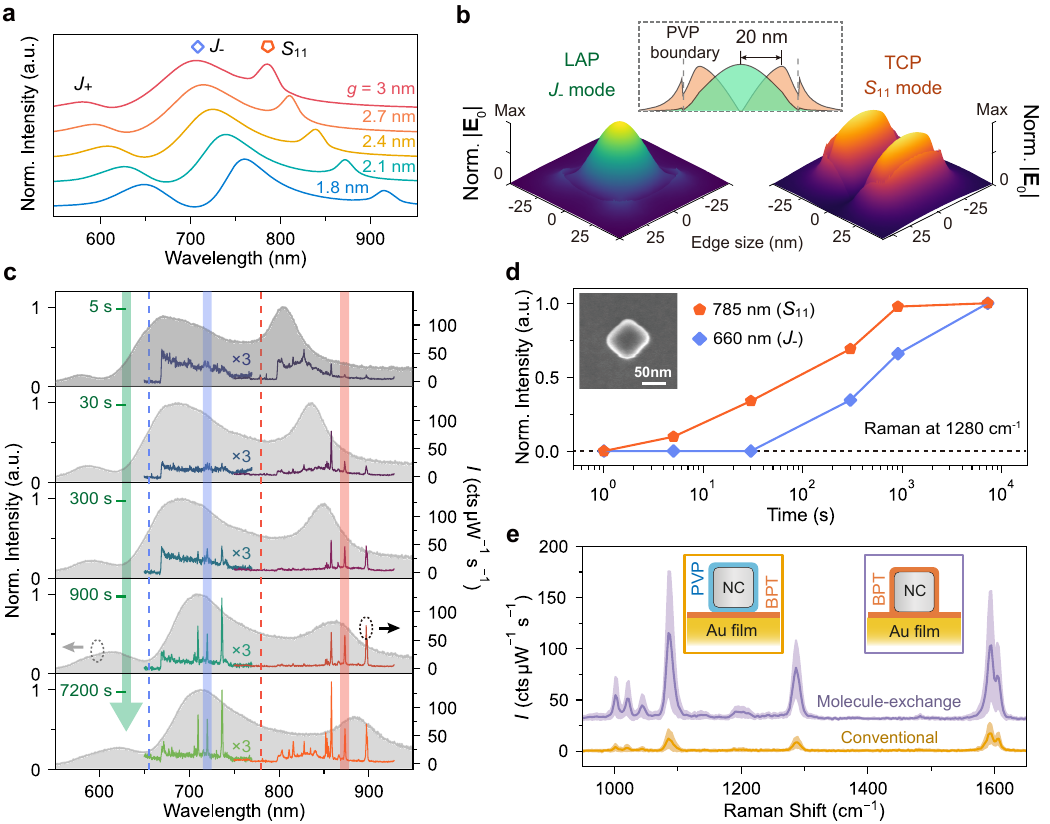}
  \caption{\textbf{Wavelength-multiplexed (WM) SERS sensing of molecular transport in a singe NCoM nanogap.}        
    (\textbf{a}) Simulated scattering spectra of a 70-nm-diameter Ag nanocube on a gold mirror with gap distance \textit{g} varied from 3 to 1.8 nm, showing three plasmon modes labeled as $J_{\text{+}}$, $J_{\text{-}}$ and $S_{\mathrm{11}}$, respectively. 
    (\textbf{b}) 3D colormap of electric field distributions of LAP $J_{\text{+}}$ and TCP $S_{\mathrm{11}}$ modes, respectively. The central panel shows the cross-sectional field intensity profiles of the LAP and TCP modes, illustrating their complementary spatial distributions and defining 20 nm spatial resolving power of the WM-SERS technique.
    (\textbf{c}) Measured dark-field scattering and dual-wavelength Raman spectra (660 and 785 nm excitation, orange and blue dashed lines) of a single PVP-capped Ag NCoM after immersion of BPT ethanol solution with duration from 5 to 7200 s (from top to bottom panels), with the SEM image shown in the inset of \textbf{d}. 
    (\textbf{d}) BPT Raman peak intensity at 1280 cm$^{\mathrm{-1}}$ as a function of immersion time under 660 nm and 785 nm excitations with Raman emission at 720.8 nm and 872.6 nm, respectively (also labelled in \textbf{c} with orange and blue areas). 
    (\textbf{e}) Statistical SERS spectra from individual BPT-spaced NCoMs using conventional and our molecule-exchange methods, respectively. The yellow (purple) line, area and inset box represents the SERS mean value, error bar and the schematics of the conventional (molecule-exchange) methods, respectively.  
    }
  \label{fig:2}
\end{figure}

To elucidate the temporal and spatial evolution of molecular infiltration within the vertical nanogap, we initiated our analysis by characterizing the scattering spectra of the previously employed NCoM nanostructure (Fig. \ref{fig:1}). 
As shown in Figs.~\ref{fig:2}a and \ref{fig:2}c, both electromagnetic simulations and experimental measurements revealed three distinct resonance peaks, labelled as well-documented \cite{chikkaraddyHowUltranarrowGap2017b} $J_{\text{+}}$, $J_{\text{-}}$ and $S_{\mathrm{11}}$ (Fig.~\ref{fig:2}a). 
Quasinormal mode analysis \cite{wu2023modal} of their near-field distributions (Fig.~\ref{fig:2}b) and corresponding polarization-dependent dark-field and SERS measurements (see Supplementary Section S5) identified the $J_{\text{+}}$ and $J_{\text{-}}$ resonances as hybridized \textit{longitudinal antenna plasmon} (LAP) mode, while the $S_{11}$ mode is a \textit{transverse cavity plasmon} (TCP) mode.
\cite{chikkaraddyHowUltranarrowGap2017b,
tserkezisHybridizationPlasmonicAntenna2015,
esteban2015morphology}
As depicted in the field profiles in Fig.~\ref{fig:2}b, the LAP ($J_{\text{-}}$) mode exhibits a single intensity maximum at the nanocube's center, whereas the TCP ($S_{11}$) mode features two distinct intensity maxima near the edges, with a node at the center. 
The distinct spectral and spatial distributions of the $J_-$ and $S_{11}$ modes enable the super-resolution mapping of molecular locations within the nanogap.
\cite{wangProbingLocationHot2014,
gopinathEngineeringMappingNanocavity2016,
griffiths_locating_2021}
The $\sim$20 nm separation between these modal maxima determines the spatial resolving power of our two-color WM-SERS technique, enabling spatial distinction at this scale or even better. 
This high-contrast spatial mapping is possible because the field maximum of the LAP mode coincides with a field null of the TCP mode, and vice versa.
Through careful geometric tuning of the NCoM (the nanocube size and the gap thickness), we deliberately align the spectral windows of $J_{-}$ ($S_{11}$) mode with the 660 (785) nm laser excitation and Stokes sideband (see details in Supplementary Section S5).
This spectral engineering enables us to selectively and independently address molecules at the nanogap's center and edges, respectively, via 660 nm and 785 nm WM-SERS measurements.


With this spatially sensitive tool established, we applied it to track the infiltration of BPT molecules into a single PVP-capped NCoM over time after immersing the sample in solutions of BPT for controlled time intervals. 
Initial evidence of molecular exchange of this NCoM is provided by dark-field spectroscopy (Fig.~\ref{fig:2}c), whose scanning electron microscopy image is shown in the inset of Fig. \ref{fig:2}d.
After sequential immersion steps, the plasmon resonances exhibit a progressive redshift, which provides a direct optical readout of a nanoscale physical change: the subsequent replacement of the thicker PVP layer ($\sim$2 nm) with a thinner layer of BPT ($\sim$1 nm), a result that is in agreement with simulations (Fig.~\ref{fig:2}a).
While this measurement confirms that exchange is occurring across the nanogap, the spatially-resolved evidence for the infiltration pathway is unlocked by the time-resolved WM-SERS spectroscopy (Fig.~\ref{fig:2}c). 
Figure 2d shows the temporal evolution of a representative BPT Raman band at 1280 cm$^{-1}$ as a function of infiltration time, appearing at 720.8 nm and 872.6 nm under 660 nm and 785 nm excitation, respectively.
The SERS signal from the edge-sensitive TCP mode (785 nm excitation) rises significantly earlier and saturates faster than the signal from the center-sensitive LAP mode (660 nm excitation). 
This distinct temporal lag is the direct signature of molecular transport, representing the finite time required for BPT molecules to infiltrate the nanochannel from the entry points at the edge to the central region.
Our observation provides unambiguous, spatially-resolved evidence for a centripetal infiltration pathway, verifying that molecules enter from the edges and progressively saturate the nanochannel towards its geometric center.

Beyond tracking the infiltration dynamics, we also evaluated the final outcome of this process by performing a statistical 785 nm SERS analysis on NCoMs after complete, saturated PVP-to-BPT exchange (Fig~\ref{fig:2}e, purple line). 
As a benchmark, we compared these results to control samples fabricated using the conventional method, where a BPT self-assembled monolayer (SAM) is formed on the Au film \textit{before} nanoparticle deposition (Fig~\ref{fig:2}e, yellow line, see more discussion in Supplementary Section S2). 
\cite{chenIntrinsicLuminescenceBlinking2021,
chenContinuouswaveFrequencyUpconversion2021a,
baumbergExtremeNanophotonicsUltrathin2019}
The comparison reveals a remarkable advantage of our strategy. 
The \textit{in-situ} molecule-exchange approach consistently yields significantly higher SERS enhancement. 
This superior performance is attributed to the formation of a cleaner and narrower nanogap. 
In the conventional scheme, residual ligands from the nanoparticle can become trapped upon the pre-formed SAM, creating an undesirably larger gap. 
In contrast, our exchange method effectively purges the initial spacer, which results in stronger and more reproducible field localization. 
While some molecules also adsorb onto the top nanoparticle surface, their SERS contribution is negligible compared to that from the extreme field confinement within the nanogap.


\subsection{Quantitative Study on Evolution Dynamics of Molecular Exchange in Plasmonic Nanogaps.}

\begin{figure}[!ht]
  \centering
  \includegraphics[width=0.9\textwidth]{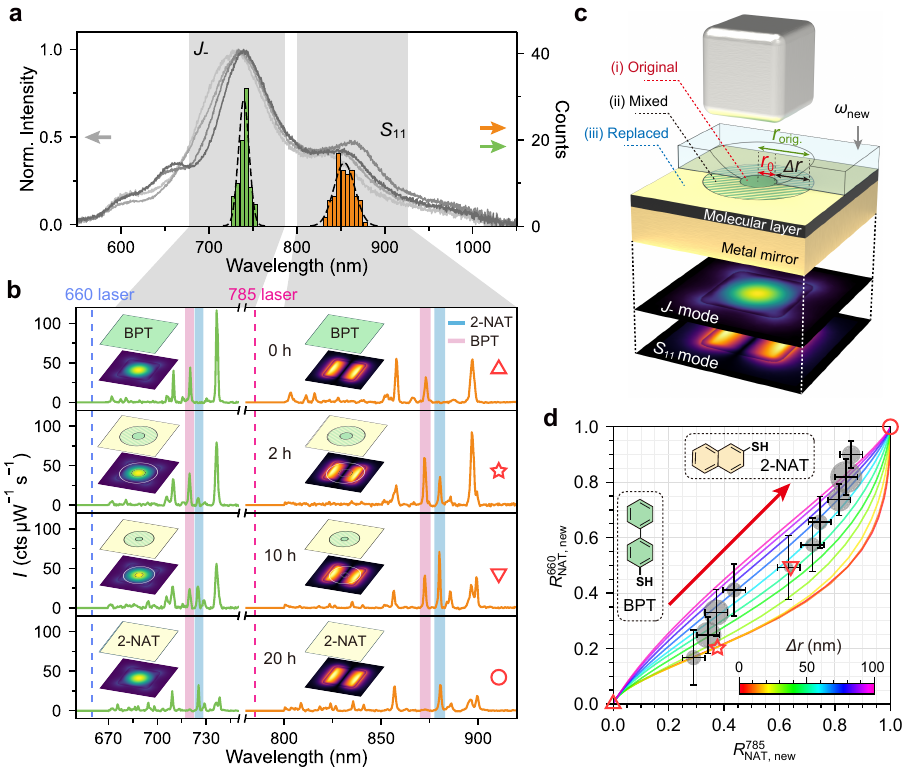}
  \captionsetup{font=small}
  \caption{\textbf{Molecular transport mechanism in plasmonic 2D nanochannels resolved by WM-SERS spectroscopy.}
    (\textbf{a}) Representative dark-field scattering spectra of NCoMs with BPT and/or 2-NAT spacer, along with the statistical distributions of resonance wavelengths for the LAP ($J_{\text{+}}$) and TCP ($S_{\mathrm{11}}$) modes from 84 NCoMs. The grey area covers the spectral windows for 660 and 785 nm SERS shown in \textbf{b}.
    (\textbf{b}) Representative WM-SERS spectra measured on individual NCoMs, representing four typical replacement stages from BPT to 2-NAT molecules, as labeled in \textbf{d}. 
    The insets illustrate the molecular spatial distribution (top plane, green: BPT, yellow: 2-NAT) within the nanogap mapped onto the near-field distributions of LAP and TCP modes (bottom planes). 
    Red and blue shaded areas highlight the characteristic Raman peaks of BPT (1280 $\text{cm}^{\mathrm{-1}}$) and 2-NAT (1380 $\text{cm}^{\mathrm{-1}}$). 
    (\textbf{c}) 3D schematics of the model for molecular exchange within a molecular-spaced NCoM nanochannel. 
    The 3D area on top of the molecular layer represents the fractional occupancy of new molecules $\omega_{\text{new}}$ (ranging from 0 to 1). The projections of $\omega_{\text{new}}$ onto the molecular layer are colored green, green-yellow striped, and yellow, respectively, representing (i) original, (ii) mixed and (iii) replaced zones.
    The corresponding WM-SERS from these three zones are mapped from the near-field distributions of the LAP and TCP modes (illustrated beneath the metal film). 
    (\textbf{d}) Correlation plot of $R_{\text{NAT}}^{\mathrm{785}}$ versus $R_{\text{NAT}}^{\mathrm{660}}$ derived from 820 individual NCoMs, where the area of the grey circles is proportional to the nanocube numbers. 
    Colored curves show the theoretically predicted relationship between $R_{\text{new}}^{\text{785}}$ and $R_{\text{new}}^{\text{660}}$ for varied $\Delta r$ of the mixed zone.
    }
  \label{fig:3}
\end{figure}

While single-particle studies provide insight into molecular exchange within nanogaps, ensemble statistics could reveal a quantitative relationship in the dynamics. 
To this end, PVP-capped NCoMs were first immersed in BPT solutions until complete replacement, yielding BPT-capped NCoMs.
Next, we induced controlled replacement of BPT spacers by immersing the structures in 2-naphthalenethiol (2-NAT) solutions for varied durations. 
This results in co-functionalized NCoMs with BPT and 2-NAT coexisting in the nanogap with different proportions and spatial distributions, where we measured over 80 NCoMs for statistical analysis.
Representative dark-field scattering spectra, and resonance wavelength statistics of $J_{-}$ mode and $S_{11}$ mode are summarized in Fig. \ref{fig:3}a.  Both peak distribution statistics were fitted with Gaussian functions. The $J_{-}$ mode exhibits a central peak position at 740~nm with a standard deviation $\sigma = 5.2$~nm 
while the $S_{11}$ mode shows a central peak at 854~nm with $\sigma = 11.8$~nm.
The spectral resonance alignment with the results in Fig. \ref{fig:2}a, showing sufficient structural homogeneity for reliable kinetic analysis.

Figure \ref{fig:3}b presents the WM-SERS spectra of four representative frames of a single NCoM after different 2-NAT immersion times (0, 2, 10, and 20 h). Insets show schematics of the molecular distributions (green: BPT, yellow: 2-NAT) overlaid with the $J_-$ and $S_{11}$ mode profiles. At 0 h, the nanogap is fully occupied by BPT, evidenced by its characteristic 1280 cm$^{-1}$ Raman peak (pink shade), while the 1380 cm$^{-1}$ peak of 2-NAT is absent. After 2 h of immersion, 2-NAT begins infiltrating from the edges and partially replaces BPT, leading to coexistence within overlapping regions (further validated in Fig. \ref{fig:3}d). In the WM-SERS spectra, the 2-NAT Raman peak at 1380 cm$^{-1}$ excited at 785 nm ($S_{11}$ mode) rises markedly, whereas the corresponding peak under 660 nm excitation ($J_{-}$ mode) shows only a slight increase. With longer immersion (10 h and 20 h), BPT is progressively replaced, and at 20 h its Raman signals vanish completely, leaving only 2-NAT. Notably, even after 10 h, despite 2-NAT reaching the nanogap center, the BPT peak under the $S_{11}$ mode persists, indicating that residual BPT molecules remain near the edges and require longer times (20 h) for full replacement. This suggests that the actual replacement dynamics are more complex than the simplified assumption of symmetric edge-infiltration, potentially involving preferred infiltration pathways that are currently non-trivial to resolve.) 
Since Raman peak intensities at different excitation wavelengths reflect the relative molecular locations, their intensity ratios can be leveraged to probe molecular dynamics in the 2D nanochannel. To achieve this, we establish a physical model and validate it through statistical analysis of experimental data (Figs. \ref{fig:3}c, d).

\subsection{A model for resolving nanogap molecular transport with WM-SERS.} We define the characteristic SERS peak intensities $I_s^\lambda$ where $s \in \{\text{NAT}, \text{BPT}\}$ denotes molecular species (2-NAT or BPT) and $\lambda \in \{660, 785\}$ nm specifies excitation wavelength. These correspond to $J_{-}$ mode signals ($\lambda = 660$ nm) originating from nanogap center and $S_{11}$-mode signals ($\lambda = 785$ nm) from the edge regions.
The absolute SERS intensity varies significantly between individual NCoMs due to subtle differences in gap size and surface morphology. 
\cite{esteban2015morphology, tserkezisHybridizationPlasmonicAntenna2015, wang2023effect}
To isolate the statistical features of molecular exchange from these inter-particle variations, we analyze the fractional SERS contribution of the 2-NAT molecules. 
This is calculated as the ratio of the 2-NAT intensity to the total SERS intensity from both molecules, rather than using the absolute SERS intensity of 2-NAT alone:
\begin{equation}
R^{\lambda}_{\text{NAT}} = \frac{I^{\lambda}_{\text{NAT}}}{I^{\lambda}_{\text{NAT}} + I^{\lambda}_{\text{BPT}}}, \quad \forall \lambda \in \{660, 785\}
\end{equation}
These ratios quantify 2-NAT coverage in the central ($R^{660}_{\text{NAT}}$) and edge ($R^{785}_{\text{NAT}}$) regions, where $R^{\lambda}_{\text{NAT}} = 0$ denotes pristine BPT coverage and $R^{\lambda}_{\text{NAT}} = 1$ indicates complete 2-NAT replacement.
As indicated by the guide-to-eyes red arrow, the trajectories in Fig. \ref{fig:3}d from lower-left (BPT-dominated) to upper-right (2-NAT-saturated) can represent the spatiotemporally resolved molecular exchange progression. The different trajectories arise from the corrections to the initial assumptions by allowing for partial molecular overlap between BPT and 2-NAT. 

As illustrated in Fig. \ref{fig:3}c, without loss of generality, we propose a simplified kinetic model in which an external new molecule (2-NAT here) infiltration proceeds toward the center, forming three concentric zones:

(i) an inner circular zone of radius $r_0$, termed "original" ($r < r_0$), containing only the original molecular spacer (BPT here, green area), where $r$ denotes the distance from the nanogap center.

(ii) an annular transition zone ($r_0 < r < r_{\text{orig.}}$) of width $\Delta r$, termed "mixed", where molecules coexist (green-yellow region). The occupancy of the new molecule described by an exponentially decaying function $\omega_{\text{new}}(r, r_\text{orig.}, \Delta r)$ and that of the original molecule by $\omega_{\text{orig.}}(r, r_\text{orig.}, \Delta r) = 1 - \omega_{\text{new}}(r, r_\text{orig.}, \Delta r)$, where $r_{\text{orig.}} = r_0 + \Delta r$ is defined as the radius of the original molecule region (see details in Supplementary Section S7).

(iii) an outer zone, termed "replaced" ($r > r_{\text{orig.}}$), fully occupied by the new molecule at saturation (2-NAT here, yellow area);

Based on well-known classical electromagnetic "$E^4$" theory of SERS, 
\cite{xuElectromagneticContributionsSinglemolecule2000}
the scattering intensity $I_s^\lambda (r_\text{orig.}, \Delta r)$ for molecular species $s$ (where $s \in \{\text{new}, \text{orig.}\}$) under excitation wavelength $\lambda$ scales as:
\begin{equation}
I_s^\lambda (r_\text{orig.}, \Delta r) \propto \left[ \iint_A |\mathbf{E}_{\lambda}(\mathbf{r})| \cdot \omega_{s}(r, \Delta r)  \,\mathrm{d}^2 r \right]^2 \cdot \left[ \iint_A |\mathbf{E}_{\lambda_{\text{em}}}(\mathbf{r})| \cdot \omega_{s}(r, \Delta r)  \,\mathrm{d}^2 r \right]^2
\end{equation}
where $\mathbf{E}_{\lambda, \lambda_{\text{em}}}(\mathbf{r})$ denotes the near field distribution of the NCoM at wavelength $\lambda$ and $\lambda_{\text{em}}$, respectively.
$\lambda_{\text{em}}$ denotes the Raman emission wavelength of the 1320 cm$^{-1}$ vibrational mode by excitation at $\lambda$ nm, with $\lambda_{\text{em}} =$ 723 nm for $\lambda = 660$ nm and $\lambda_{\text{em}} =$  875 nm for $\lambda =$ 785 nm (See detail discussion in Supplementary Section S7).
$A$ is an integration area with a radius set to 100 nm to cover the whole near field distribution.
In comparison with $R^{\lambda}_{\text{NAT}}$, the theoretical normalized SERS intensity ratios of the new molecules are thus obtained as:
\begin{align}
R_{\text{new}}^{\lambda}  (r_\text{orig.}, \Delta r) = \frac{I_{\text{new}}^{\lambda}}{I_{\text{new}}^{\lambda} + I_{\text{orig.}}^{\lambda}}, \quad \lambda \in \{660, 785\} \
\end{align}

Solid colored lines in Fig.~\ref{fig:3}d represents the computed function between $R_{\text{new}}^{\text{660}}  (r_\text{orig.}, \Delta r)$ and $R_{\text{new}}^{\text{785}}  (r_\text{orig.}, \Delta r)$, where $r_{\text{orig.}}$ spanned from 0 to 100 to cover the overall molecular exchange process. As shown in the different curves in Fig.~\ref{fig:3}d, the process are also dependent on the degrees of molecular overlap $\Delta r$. The Raman intensities of time-resolved data from the four frames of a single NCoM in Fig.~\ref{fig:3}b are calculated and overlaid in Fig.~\ref{fig:3}d as red open markers. Their positions (from left to right on the map) reflect the logical time sequence from 0 to 20 h, indicating that the 2D map in Fig.~\ref{fig:3}d captures both the temporal and spatial distribution of the molecules.

The parameter $\Delta r$ quantifies the spatial extent of molecular mixing near the interaction frontier. 
At the theoretical limit, $\Delta r \to \infty$, the model describes a complete lack of spatial confinement or directional transport. 
This scenario implies that the exchange is not limited by infiltration from the edge inwards; rather, it occurs uniformly and simultaneously across the entire nanogap area. 
This scenario is physically equivalent to the process of exchange occurring on an open gold surface, 
\cite{villarrealNanoscaleSurfaceCurvature2017, tukovaShapeInducedVariationsAromatic2023}
where different types of molecules are homogeneously mixed without anisotropic selectivity and direction of interaction caused by the geometric confinement.
This stands in stark contrast to the $\Delta r = 0$ case, which represents a perfectly sharp, inward-propagating replacement front without co-existence of two types of molecules.

We measured a total of 830 nanoparticles and extracted the characteristic peaks of 2-NAT (1380 cm$^{-1}$) and BPT (1280 cm$^{-1}$) from each particle. Based on the ratio of their peak intensities, the nanoparticles were classified into ten groups. The size of the gray disk represents the number of samples in each group, while the position and error bar indicate the mean and variation, respectively. Interestingly, the “centers of mass” of generally binned data remain well within the region predicted by the analytical solutions.  
By comparing the experimental data with our analytical solution, we find that the transition zone width, $\Delta r$, spans a range from 0 to 100 nm, with an average value of approximately 50 nm. 
This significant average value, which exceeds half the edge length of the 70 nm nanocubes, leads to a crucial insight: for a substantial fraction of NCoMs, the purely "original" central region (zone (i) in Fig.~\ref{fig:3}c) virtually does not exist throughout the process. This implies that infiltrating molecules can penetrate to the geometric center of the nanogap from the earliest stages of the replacement. 
It occurs even in thiol-based, SAM-spaced nanogaps ($\sim$1-1.3 nm) comparable to the molecular dimensions (BPT versus 2-NAT), challenging the intuitive assumption that the nanoparticle would physically block molecular transport and seal the central region. 
The experimental errors reflect a degree of randomness in the infiltration pathways, which may deviate from the idealized concentric model due to local nano-environmental factors. 
\cite{wang2023effect}
Nevertheless, the model effectively captures the overall experimental trend.


\subsection{Mapping Molecular Infiltration Dynamics in Extreme-Aspect-Ratio Nanochannels.} 
\begin{figure}[!ht]
  \centering
  \includegraphics[width=1\textwidth]{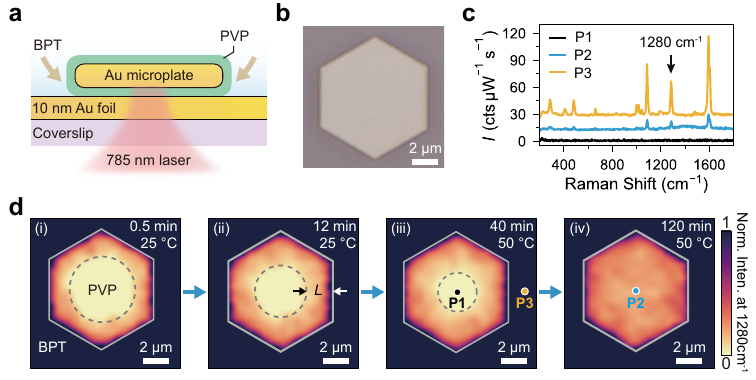}
  \caption{\textbf{Micron-scale molecular infiltration and exchange dynamics.}
    (\textbf{a}) Cross-sectional schematic of the molecular transport experiment in a microplate-on-foil (MPoF) nanochannel with micro length. The sample consists of a PVP-capped microplate placed on a 10-nm-thick Au foil, forming a $\sim$2-nm-thick gap. The MPoF was then immersed in BPT solution for varying durations to probe molecular exchange process via SERS mapping.
    (\textbf{b}) Bright field image of a representative MPoF structure. 
    (\textbf{c}) SERS spectra of the MPoF collected at the position before (P1, black dot in \textbf{d} (iii)) and after (P2, blue dot in \textbf{d} (iv)) BPT replacement, respectively, compared with that at the BPT-adsorbed Au foil region (P3, yellow dot in \textbf{d} (iii)). For clarity, the blue and yellow spectra are vertically offset.
    (\textbf{d}) Raman intensity maps at 1280 cm$^{-1}$ (785 nm excitation) for the MPoF in \textbf{b} after immersion in BPT solution for (i) 0.5 min, $25 ^\circ$C, (ii) 12 min, $25 ^\circ$C, (iii) 40 min, $50^\circ$C, and (iv) 120 min, $50 ^\circ$C, respectively.
    The dashed black circle and gray boxes indicate the original PVP-spacer region and boundary of the microplate, respectively. $L$ denotes the length of the microplate region where molecular infiltration has occurred.
    }
  \label{fig:4}
\end{figure}

Building on our nanoscale model of molecular transport, we next sought to directly visualize this process in real space. To this end, we scaled up the NPoM-like architecture to the micron level by engineering nanochannels with extreme aspect ratios through a microplate-on-foil (MPoF) architecture (Fig.~\ref{fig:4}a). 
This configuration was realized by depositing a $\simeq2$ nm PVP-capped gold microplate (edge dimensions: 5--50 $\mu$m; thickness: 50--100 nm) onto a 10-nm-thick rough Au foil thermally evaporated on a glass coverslip (see Supplementary Section S1). The MPoF platform generates micron-long plasmonic nanocavities with a gap around 2 nm (Fig.~\ref{fig:4}a). The bright field image of the microplate was shown in Fig.~\ref{fig:4}b.

We chose a 10-nm-thick Au foil because it both transmits light and forms a quasi-uniform nanogap as plasmonic hotspots \cite{chikkaraddy2021accessing}, enabling spatially resolved SERS of molecular transport dynamics inside the nanochannel (Fig.~\ref{fig:4}a). We investigated infiltration kinetics by immersing PVP-covered MPoF specimens in BPT solution while modulating incubation time and temperature (to accelerate the exchange). Hyperspectral SERS mapping across entire microplates revealed the spectral and spatiotemporal progression of BPT replacement of the PVP spacer layer. Representative spectra as well as spatial maps of the 1280 cm$^{-1}$ Raman band are presented in Fig.~\ref{fig:4}c and \ref{fig:4}d, respectively. 

Molecular infiltration and exchange are directly visualized via mapping of BPT’s characteristic SERS peak in Fig. \ref{fig:4}d. The infiltration patterns (0.5 to 40 min incubation, Figs.~\ref{fig:4}d(i-iii)) reveal three distinct and progressively evolving zones: (1) uniform SERS signals outside the microplate boundary (its representative spectrum shown as yellow line in Fig.~\ref{fig:4}c, position P3), corresponding to BPT monolayers adsorbed on the foil’s plasmonic hotspots induced by its roughness; (2) a central signal-deficient circular region (black line in Fig.~\ref{fig:4}c, position P1), indicating unmodified PVP spacers; and (3) a region near the microplate edge showing weaker but discernible BPT signatures, confirming nanochannel infiltration over distances of $L \sim 2~\, \mu$m as shown in Fig. \ref{fig:4}d(iii). With increasing incubation time, the central PVP region progressively disappears (Fig. \ref{fig:4}d(iv)), while BPT signals gradually emerge at the nanogap center (shown by the comparison of spectra in Fig. \ref{fig:4}c at the position P1 and P2 in Fig. \ref{fig:4}d(iii,iv)). These provide a direct visualization of molecular propagation and exchange.

This real-space observation of micron-scale molecular transport within $\sim$2-nm-thick 2D channels validates our predictive infiltration model shown in the last sections. Critically, we demonstrate unprecedentedly long molecular infiltration lengths in extreme-confinement environments ($\sim$10$^3$ aspect ratio). 
Our findings are particularly significant as molecular transport dynamics within these extreme-aspect-ratio 2D nanochannels—where confinement is comparable to the molecular scale—are expected to differ fundamentally from behavior in larger spaces. 
While other platforms like plasmonic nanopores exist, their channel dimensions are often significantly larger than the target molecules, leading to stochastic measurement challenges. 
\cite{spitzbergPlasmonicNanoporeBiosensorsSuperior2019}
In contrast, our system provides a versatile platform for simultaneous optical characterization under these extreme, molecular-scale confinement conditions. 

\subsection{Real-time and Ultrasensitive Molecular Sensing Platform.}
\begin{figure}[!ht]
  \centering
  \captionsetup{font=small}
  \includegraphics[width=1\textwidth]{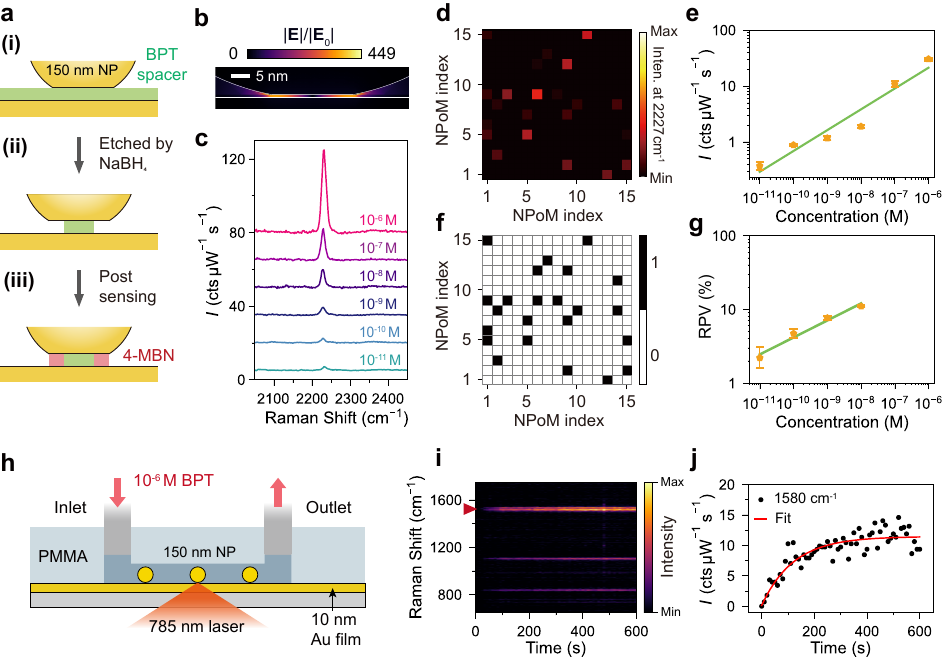}
  \caption{\textbf{\textit{In-situ} ultrasensitive and real-time molecular post-sensing via a single NPoM plasmonic nanogap.} 
    (\textbf{a}) Schematic of the selective nanogap engineering process for molecular post-sensing: (i) A BPT-spaced 150-nm-diameter Au NPoM is formed via the molecular exchange method; 
    (ii) Partial etching of edge BPT molecules using NaBH$_{4}$ solution preserves central BPT to maintain nanochannel integrity; 
    (iii) Subsequent immersion in analyte (4-MBN) solution enables analyte adsorption into the etched nanogap regions.
    (\textbf{b}) Near-field distribution (at 880 nm) of the 150 nm NPoM, providing field enhancement over 449 times.
    (\textbf{c}) Representative 785-nm SERS spectra showing the characteristic 4-MBN peak at 2227 cm$^{-1}$ across concentrations varied from 10$^{-6}$ to 10$^{-11}$ M. The spectra are vertically offset for clarity.
    (\textbf{d}) Raman intensity map (2227 cm$^{-1}$) of 225 individual NPoM structures at 10$^{-8}$ M 4-MBN, arranged in a 15$\times$15 grid.
    (\textbf{e}) Statistical SERS intensity at 2227 cm$^{-1}$ of the NPoM as 4-MBN concentrations varied from 10$^{-6}$ to 10$^{-11}$ M. Green line denotes a linear fit. 
    (\textbf{f}) Digital SERS map derived from (\textbf{d}) by applying an intensity threshold (see Supplymentary Section 6). 
    (\textbf{g}) Ratio of positive voxels (RPV) based on digital SERS analysis (\textbf{f}) versus 4-MBN concentration. The green line denotes a linear fit. 
    (\textbf{h}) Microfluidic experimental setup for real-time SERS: A CTAC-capped Au nanoparticle (150 nm diameter) on Au foil (10 nm thick) forms a nanoparticle-on-foil (NPoF) nanogap.
    10$^{-6}$ M BPT solution is flowed through the microchannel while acquiring 785-nm SERS spectra from the coverslip side.
    (\textbf{i}) Time-dependent SERS map during microfluidic perfusion over 600 s. 
    (\textbf{j}) Time-dependent SERS intensity (BPT peak at 1580 cm$^{-1}$) during microfluidic perfusion, which is fitted by an exponential function based on the Langmuir kinetic model.}
  \label{fig:5}
\end{figure}

Having established the molecular exchange mechanism within plasmonic 2D nanochannels, we now evaluate the post-sensing capabilities of these easy-fabricated NPoM nanostructures. An effective sensing platform was prepared by firstly assembling 150 nm--diameter CTAC-capped gold nanoparticles onto gold films, followed by immersion the sample in BPT solution to achieve saturated spacer replacement. 
To create active binding sites for subsequent analyte detection, the BPT-spaced NPoMs were briefly treated with NaBH$_{4}$ solution (procedures shown in Fig.~\ref{fig:5}a). 
This partial etching process is expected to initiate at the more accessible peripheral regions of the nanogap, creating a higher density of vacant sites near the edges for analyte infiltration.
This edge region supports well-defined hot-spot regions with field enhancement over 400 folds for molecular sensing (Fig.~\ref{fig:5}b). 
We term it post-sensing because these NaBH$_4$-etched open 2D channels (chambers) with nanogap-hotspots can serve as a general SERS substrate, capable of detecting molecules deposited afterward (see details in Supplementary Section S4). 

To evaluate the sensing performance, the modified NPoMs (Fig.~\ref{fig:5}a(ii)) were incubated in 4-MBN solutions of varying concentrations, allowing the analyte to occupy the nanogap regions previously etched by NaBH$_4$. 
The representative SERS spectra are shown in Fig.~\ref{fig:5}c and Supplementary Fig. S7. 
The characteristic nitrile stretch of 4-MBN at 2227 cm$^{-1}$ is spectrally well-separated from the Raman bands of the BPT spacer, enabling unambiguous signal characterization. 
For each concentration (from $10^{-6}$ M to $10^{-11}$ M), we measured 225 individual NPoMs (see details in Supplementary Section S6). 
To visualize the statistical variations, the SERS intensities of the characteristic peak were arranged into a 15 $\times$ 15 grid map, as exemplified for $10^{-8}$ M in Fig.~\ref{fig:5}d. 
By averaging the intensities for each concentration, we plotted the SERS intensity as a function of molecular concentration (Fig. \ref{fig:5}e), which displays a generally monotonic, quasi-linear trend.
Recent advances in SERS show that the linear relationship between signal intensity and concentration breaks down at low concentrations; we also employed a digital SERS approach. 
\cite{biDigitalColloidenhancedRaman2024,
zhangStatisticalRouteRobust2024}
This technique more accurately describes analyte presence by quantifying the probability of discrete molecular adsorption events ("on" vs. "off"). 
To implement this, we set an intensity threshold for the 4-MBN peak: signals above the threshold were assigned a value of 1, and those below were assigned 0 (see details in Supplementary Section S6). 
This converted the analog intensity map (Fig.~\ref{fig:5}d) into a binary map of detection events (Fig.~\ref{fig:5}f). Plotting the resulting Ratio of positive voxels (RPV) against concentration (Fig.~\ref{fig:5}g) revealed an improved linear dynamic range. This digital analysis further confirms our ability to detect 4-MBN down to $10^{-11}$ M, establishing the ultrasensitive sensing of our platform for non-resonant small molecules toward single-molecule detection limits.
\cite{dealbuquerqueDigitalProtocolChemical2018, biDigitalColloidenhancedRaman2024}

Real-time molecular detection is an essential capability in Raman spectroscopic applications. We integrated a nanoparticle-on-foil (NPoF) architecture with a microfluidic device, as depicted in Fig.~\ref{fig:5}h. 
The NPoF platform, comprising 150-nm-diameter CTAC-capped gold nanoparticles on a 10-nm-thick Au foil, was specifically designed to facilitate both 785 nm laser excitation and \textit{in situ} single-particle SERS signal collection. We perfused a $10^{-6}$ M BPT solution through the microfluidic channel, allowing it to flow over and interact with the NPoF structures (Fig.~\ref{fig:5}h). 
Simultaneously, we performed continuous \textit{in situ} SERS measurements on a single NPoF to monitor the process in real time, as shown in Fig.~\ref{fig:5}i. 
It is clear that as the perfusion time increases, every distinct Raman peak becomes more prominent. We extracted the intensity of the BPT characteristic peak at 1580 cm$^{-1}$ (marked by the red triangles) and plotted it as a function of perfusion time in Fig. \ref{fig:5}j. 
The fitting of temporal evolution follows a clear exponential saturation curve, which is consistent with Langmuir kinetic model of molecular infiltration, replacement, and eventual saturation (see details in Supplementary Section S8). 
\cite{tukovaShapeInducedVariationsAromatic2023}
This experiment not only provides further validation for our analysis of molecular dynamics above, but also demonstrates that NPoM-like systems can effectively serve as an all-in-one "transport-and-probe" platform based on nanoconfined 2D channels.



\section{Discussion}
Our work fundamentally unlocks a post-sensing paradigm for bottom-up plasmonic nanogap architectures. 
While bottom-up fabrication has been superior to top-down methods in creating reliable, sub-nanometer gaps with exceptional field confinement, its potential has been constrained by the perceived "closed" nature of the nanojunctions. 
By demonstrating that NPoM-like gaps function as open nanochannels accessible via molecular transport and exchange, we bridge the gap between high-performance plasmonics and dynamic, \textit{in situ} analysis.
While our work focused on individual NPoMs to elucidate the underlying mechanism, this newly established open-channel functionality inherently benefits from the well-known scalability of the NPoM platform, which offers near-100\% yield over macroscopic areas at a fraction of the cost of top-down methods.
\cite{liPlasmonicParticleonfilmNanocavities2018,
baumbergExtremeNanophotonicsUltrathin2019}

Our experimental validation intentionally focused on the simplest ligand-spaced NPoM configuration to reveal its counterintuitive
\cite{wang2020fundamental}
yet powerful capabilities. 
This basic architecture offers a dynamic nanochannel where initial weak-binding ligands can be readily replaced by most small molecules. 
However, even when a dense self-assembled monolayer formed by thiols occupies the gap, the channel remains accessible. 
Subsequent exchange with molecules possessing smaller binding energy is not prevented, but simply becomes less kinetically favorable. 
For example, our results show that MB can still infiltrate the gap to replace the pre-adsorbed BPT monolayer (Supplementary Fig. S12). 
This finding challenges the notion of a sealed channel after strong ligand adsorption and reinforces the high versatility of our platform.
On the other hand, the partial etching strategy demonstrated in Fig.~\ref{fig:5}a represents a straightforward solution, creating active sensing sites while preserving the structural integrity of the nanogap. 
This simple, low-cost approach is sufficient and ideal for many applications, where, e.g., the post-adsorption of analytes is ideal for differential measurements by tracking weak signal responses. 
\cite{taylor2017single,
shiActiveDeliverySingle2018,
chenProbingSubpicometerVertical2018}
In chiroptical sensing, 
\cite{warningNanophotonicApproachesChirality2021a}
this would allow the intrinsic chiral response of the plasmonic structure to be measured first, followed by the introduction of chiral molecules, effectively decoupling the analyte's signal from the system's background chirality. 
This concept extends to tracking chemical reactions and energy transfer,
\cite{choiMetalCatalyzedChemicalReaction2016,
oksenbergEnergyresolvedPlasmonicChemistry2021a,
zhanPlasmonicNanoreactorsRegulating2021,
lee2021chemical,
huAlchemicallyglazedPlasmonicNanocavities2025}
or any externally stimulated transformation that is sensitive to the plasmonic extreme local field. 
In studies of photocatalysis that prefer a well-defined metallic surface, 
\cite{ding2014ligand}
the molecular exchange process we demonstrated could be an intrinsic and necessary step to remove the original ligands that would otherwise impede the reaction.

Looking ahead, the NPoM transport-and-probe platform offers various possibilities, from accessible, cost-effective solutions to highly engineered, high-performance systems. 
For applications demanding greater specification and reproducibility, the simple ligand spacer could be replaced with more sophisticated scaffolds such as MOFs,
\cite{liuMetalOrganicFrameworkEnabled2024}
host-guest complexes, 
\cite{kimSmartSERSHot2018}
artificial channels, 
\cite{shenArtificialChannelsConfined2021}
or using nanoparticles with concave surfaces.
\cite{zhangNobleMetalNanocrystalsConcave2012}
This path, however, presents a trade-off between performance and complexity, requiring careful consideration based on the specific application. 
For more ambitious goals, such as achieving fast mass transport, the platform must be engineered to combine the performance metrics of plasmonic nanopores 
\cite{spitzbergPlasmonicNanoporeBiosensorsSuperior2019,
garoliPlasmonicNanoporesSingleMolecule2019}
with the unique geometric advantages of the 2D nanochannel—namely, extended analyte dwell times and the optimal field enhancement of the NPoM nanogaps. 
The future development of this platform can draw inspiration from the capabilities established in these related fields, aiming to track the transport and reactions of single molecules with ever-higher spatiotemporal resolution. 
This could be achieved by integrating advanced super-resolution techniques 
\cite{brasseletSinglemoleculeOrientationLocalization2025}
in terms of spatial localization 
\cite{leeImagingNanoscaleLight2020,
willetsSuperResolutionSurfaceEnhancedRaman2024}
and molecular orientation, 
\cite{huijbenPointSpreadFunctionDeformations2024}
which would complement the WM-SERS method used here.
Furthermore, analogous to established strategies for dynamic control in the broader field of 2D nanochannels, 
\cite{zhouElectricallyControlledWater2018,
zhangControllableIonTransport2019}
active tuning of the NPoM gap size could enable molecular sieving, adding a layer of selectivity to the ultrasensitive detection.
Finally, the microscopic details of the surface chemistry in such a nanochannel—including molecular orientation within the gap and preferential binding sites—warrant further investigation to build a complete picture of this complex nanoscale environment.

In conclusion, this study establishes molecular exchange as a transformative methodology that redefines ligand-capped plasmonic nanocavities as open-channel, all-in-one "transport-and-probe" platforms. 
Our approach preserves the defining advantages of bottom-up assembly with sub-nanometer precision and reproducible electromagnetic hotspots, while overcoming the limitation of analyte inaccessibility. 
We have demonstrated that this architecture enables post-fabrication molecular infiltration using WM-SERS, achieving near-single-molecule sensitivity ($10^{-11}$ M) for nonresonant analytes and real-time monitoring in microfluidic environments. 
By visualizing molecular transport over unprecedented length scales (> 5 $\mu$m) within tightly confined ($\sim$2 nm) channels with extreme aspect ratios (>10$^3$), we resolve important questions regarding molecular ingress in vertically confined ultrasmall nanogaps. 
The proposed centripetal replacement model, which challenges classical infiltration expectations, provides a new framework for understanding these processes. 
Fundamentally, this work establishes open-channel NPoM-like structures as a simple, scalable, and universal architecture for advancing \textit{in situ} ultrasensitive science and exploring novel transport phenomena at the nanoscale.

\section*{Supplementary Materials}
Supplementary Materials for this paper include the following sections:
\begin{itemize}
    \item[S1.] Methods
    \item[S2.] Comparative analysis of molecule-exchange vs. conventional fabrication for NPoMs
    \item[S3.] The role of spacer molecules in enabling nanochannel accessibility
    \item[S4.] Engineering open sensing sites via partial etching with NaBH$_4$
    \item[S5.] Gap mode analysis via polarization-dependent SERS and DF spectroscopy
    \item[S6.] Statistical framework for digital SERS analysis at low concentrations
    \item[S7.] Theoretical model for spatially resolved molecular replacement
    \item[S8.] Kinetic analysis of sequential molecular exchange in the plasmonic nanogap
\end{itemize}

\section{Acknowledgments}
This work was supported by the National Key Research and Development Program of China (Grant No. 2024YFA1409902) and the National Natural Science Foundation of China (Grant No. 62475071 and 52488301).

\section{Conflict of interests}
The authors declare no competing interests.


\section{Data availability}
Source data, codes, and models are available at{https://doi.org}. All other data that support the plots within this paper and other findings of this study are available from the corresponding author upon reasonable request.

\clearpage
\bibliography{GapSensing_20250317}

\end{document}